%Paper: astro-ph/9301003
%From: "Antonio Lanza, Sissa - Trieste" <lanza@tsmi19.sissa.it>
%Date: Fri, 08 Jan 1993 12:10:57 +0100

%%%%%%%%%%%%%%%%%%%%%%%%%%Text start here%%%%%%%%%%%%%%%%%%%%%%
\magnification =1200
\font\tbf=cmbx10 scaled \magstep2
\baselineskip=12pt
\parskip=0.2cm
\parindent=1cm
\def\ref{\par\noindent\hangindent=1 truecm}
\def\simless{\mathbin{\lower 3pt\hbox
   {$\rlap{\raise 5pt\hbox{$\char'074$}}\mathchar"7218$}}}
\def\simgreat{\mathbin{\lower 3pt\hbox
   {$\rlap{\raise 5pt\hbox{$\char'076$}}\mathchar"7218$}}}
\null
%%%%%%%%%%%%%%%%%%%%%%%%%%%%%%%%%%%%%%%%%%%%%%%%%%%%%%%%%%%%%%%%%%
\rightline{SISSA - Ref. 220/92/A}
\vskip 2truecm
\centerline {\tbf Similarity of the variability patterns}
\vskip 0.02truecm
\centerline {\tbf in the Exosat and Ginga folded light curves}
\vskip 0.02truecm
\centerline {\tbf of the Seyfert galaxy NGC 6814}
\vskip 1.2truecm
\centerline {M.A. Abramowicz$^{1,~2, ~3}$, G.
Bao$^{1, ~5}$,
V. Karas$^{1, ~2, ~4}$,
A. Lanza$^1$}
\vskip 0.25truecm
\centerline{$^1$SISSA, Via Beirut 4, I-34014 Trieste, Italy}
\centerline{$^2$Nordita, Blegdamsvej 17, DK-2100 K{\o}benhavn {\O},
     Denmark}
\centerline{$^3$ICTP, Strada Costiera 11, I-34014 Trieste, Italy}
\centerline{$^4$ Astronomical Institute, Charles University,
{\u S}v\'edsk\'a 8,}
\centerline{CS-150 00 Prague, Czechoslovakia}
\centerline{$^5$ Department of
Physics, The University of Trondheim,}
\centerline{ N--7055, Dragvoll, Norway}
%%%%%%%%%%%%%%%%%%%%%%%%%%%%%%%%%%%%%%%%%%%%%%%%%%%%%%%%%%%%%%%%%%
\vskip 3.0truecm
%\vfill\eject
\centerline {\bf ABSTRACT}
\vskip 0.4truecm
\noindent
The Seyfert galaxy NGC 6814 is known to show periodic variation
of its X-ray luminosity. We found that the sequences of peaks
(variability patterns) in the folded X-ray light curves
constructed from the {\it Exosat} and {\it Ginga} data are
remarkably similar when one ignores amplitudes of the peaks and
considers only their phases.  The stable pattern consists of
five peaks which are
present in the both curves. The phases of the corresponding
peaks coincide with an accuracy of about 10 degrees.  The
probability that this coincidence occurs by chance is less than
about 1\% according to the most conservative estimate. The
observed stable pattern of peaks may be produced by a stable
distribution of ``bright spots'' on the accretion disk surface,
{\it e.g.} by strong vortices or magnetic flux tubes.
\vskip 2cm
\centerline {\it To appear in Astronomy \& Astrophysics }
%%%%%%%%%%%%%%%%%%%%%%%%%%%%%%%%%%%%%%%%%%%%%%%%%%%%%%%%%%%%%%%%%%%%%%%%%%%%%%%
\vfill
\eject
%%%%%%%%%%%%%%%%%%%%%%%%%%%%%%%%%%%%%%%%%%%%%%%%%%%%%%%%%%%%%%%%%%%%%%%%%%%%%%%

%%%%%%%%%%%%%%%%%%%%%%%%%%%%%%%%%%%%%%%%%%%%%%%%%%%%%%%%%%%%%%%%%%%%%%%%%%%%%%%
\centerline {\bf 1. Introduction}
\vskip 0.4truecm
%%%%%%%%%%%%%%%%%%%%%%%%%%%%%%%%%%%%%%%%%%%%%%%%%%%%%%%%%%%%%%%%%%%%%%%%%%%%%%%

The Seyfert galaxy NGC 6814 shows a very stable periodic behaviour in the form
of repeated X-ray flares with a period $P_0 = 12~200$ sec in
both the {\it Exosat} data (Mittaz \& Branduardi-Raymont 1989;
Fiore {\it et al.} 1992a, b) and the {\it Ginga} data (Done
{\it et al.} 1992a, b). The stability of the period estimated by
Fiore {\it et al.} is ${\dot P}_0 \simless 5\times 10^{-6}$.
Theoretical models which had been suggested by several authors to explain the
observed periodic flaring have been critically discussed by
Abramowicz (1992), and the main points of his discussion were
summarized in a recent review article by Wallinder {\it et al.}
(1992).

In this paper we describe a strong observational constraint
for possible theoretical models which we have found by comparing the
folded light curves constructed from the {\it Exosat} and {\it Ginga} data.
Figures 1 and 2 show the NGC 6814 light curves obtained by {\it
Exosat} (Fiore {\it et al.} 1992b) and {\it Ginga} (Done {\it et
al.} 1992b), both folded on the period $P_0 = 12~200$ sec. The
{\it Exosat} folded light curve clearly shows one major flare
and less clearly several smaller flares, while the {\it Ginga}
light curve shows three major flares and several smaller ones.
Although the first impression may be that these curves are very
different, a closer look reveals a striking similarity between
them: if one ignores amplitudes of the flares and considers only their
phases, one finds that
the phases of the corresponding five {\it Exosat} and {\it Ginga} flares
differs by less than 10 degrees. Probability that this coincidence
may occur by a pure chance is very small and thus it is quite
possible that we do observe a pattern of five flares which have not
changed their phases during the six years between the {\it Exosat} and
{\it Ginga} observations.

%%%%%%%%%%%%%%%%%%%%%%%%%%%%%%%%%%%%%%%%%%%%%%%%%%%%%%%%%%%%%%%%%%%%%%%%%%%%%%%
\vskip 0.4truecm
\centerline {\bf 2. Fitting the variability patterns to the bright spots
model}
\vskip 0.4truecm
%%%%%%%%%%%%%%%%%%%%%%%%%%%%%%%%%%%%%%%%%%%%%%%%%%%%%%%%%%%%%%%%%%%%%%%%%%%%%%%

Abramowicz {\it et al.} (1991, 1992a, b, c) have
proposed that
the observed periodic variability of NGC 6814 is due to the orbital
motion of a pattern of a few bright spots located somewhere on the
central part
of an accretion disk. Modulation of the observed intensity
is due to the relativistic Doppler effect, gravitational lensing
and occultations by the outer parts of the disk. The
gravitational lensing occurs when one of the bright spots moves,
with respect to the observer, almost exactly behind the central
black hole and this situation only happens for accretion disks with a
nearly edge-on orientation.

For an assumed spot pattern (phase $\varphi_j$, size
$\Delta\varphi_j$, intensity $I_j$ of each individual spot, {\it
i.e.} for $j=1,...,N$ where $N$ is the number of spots) and for
an assumed set of orbital parameters (mass of the central black
hole $M$, orbital radius $r$, inclination of the axis of the
accretion disk with respect to the line of sight $\theta$) one
computes the exact shape of the model light curve. By fitting it
to the observed light curve, one deduces the values of the
parameters describing the spot pattern and the orbit.  Technical
details of this procedure (which includes all general
relativistic effects with no approximations) have been described
by Abramowicz {\it et al.} (1992b), Zhang and Bao (1991), Bao (1992),
Karas and Bao (1992), and Bao and Stuchl{\'\i}k (1992).

The existence of bright spots is not postulated {\it ad hoc} to
explain the variability of NGC 6814, but it was suggested a few
years earlier by Abramowicz {\it et al.} (1989) in connection
with a more typical, noise-like, X-ray variability found for
most AGN. Typically, the X-ray variability of AGN
in the range of time scales between $10^3$ sec and $10^5$
sec is featureless and shows no preferred frequency, $f$.
Its power can be approximated by $1/f^{\beta}$ noise, with $1
\le \beta \le 2$.  Abramowicz {\it et al.} (1989, 1991)
demonstrated that this can be explained by orbital motion of
several hundred of small spots at a range of radii on the
accretion disk surface.  The idea of bright spots on accretion
disks in active nuclei is now gaining strong observational
support (see {\it e.g.} Veilleux \& Zheng 1991; Zheng {\it et
al.} 1991; Miller {\it et al.} 1992; Witta {\it et al.}, 1991,
1992; Dultzin-Hacyan {\it et.  al}  1992). In this context NGC
6814 is unique because in addition to several hundreds small
spots which produce the observed (and typical) $1/f$ noise, it
also has a few very strong spots which produce the observed
periodic signal (unique to this source).

According to the bright spots model, the brightness of each
individual spot may change from period to period, but the
positions (orbital phases) of the strong spots should remain
fairly constant, although not exactly fixed, as they should
reflect the ${\dot P}_0 \simless5\times 10^{-6}$ period stability.
Thus, there should be the same variability pattern defined by
the strong spots in the {\it Exosat} and {\it Ginga} light
curves --- repeated major flares should occur at almost the same
orbital phases independently of their varying amplitudes. As a
consequence, phase differences between the major flares should
be almost constant in time, {\it i.e} they should be constant
with a similar accuracy as the period is constant in time.

\smallskip
{\it i) The fits to Exosat and Ginga folded light curves.}
\smallskip

We have fitted to both {\it Exosat} and {\it Ginga} data several
theoretical light curves calculated from the bright spots model.
The quality of the fit was judged by the $\chi^2$ test
calculated according to the formula,
%%%%%%%%%%%%%%%%%%%%%%%%%%%%%%%%%%%%%%%%%%%%%%%%%%%%%%%%%%%%%%%%%%%%%%%%%%%
$$ \chi^2 = {1\over {k - n}} \sum_{i=1}^{k}\left ( {{I_{\rm
O}({\varphi_i}) - I_{\rm M}({\varphi_i})}\over \sigma_{\rm
O}(\varphi_i)}\right)^2.
\eqno (2.1)$$
%%%%%%%%%%%%%%%%%%%%%%%%%%%%%%%%%%%%%%%%%%%%%%%%%%%%%%%%%%%%%%%%%%%%%%%%%%%
Here $\varphi_i$, $I_{\rm O}(\varphi_i)$, $\sigma_{\rm
O}(\varphi_i)$ are the phases, intensities (measured in counts
per second), and errors of the observed $k$ data points, $I_{\rm
M}(\varphi_i)$ are the intensities calculated from the model,
and $n = 3\times N + 4$, where $N$ is the number of the spots in
the model, is the number of the free parameters.

First we found that if the orbital parameters are the same for
both {\it Exosat} and {\it Ginga}, then the best fit model
gives,
%%%%%%%%%%%%%%%%%%%%%%%%%%%%%%%%%%%%%%%%%%%%%%%%%%%%%%%%%%%%%%%%%%%%%%%%%%%%
$$ M = 9\times 10^6 M_{\odot},~~~~~~
r = 6  r_G,~~~~~~\theta = 85^o.
\eqno (2.2) $$
%%%%%%%%%%%%%%%%%%%%%%%%%%%%%%%%%%%%%%%%%%%%%%%%%%%%%%%%%%%%%%%%%%%%%%%%%%%%%

We then made fits with the orbital parameters (2.2) fixed, but with
several different numbers of spots and several different spot
patterns.  The best fits (minima of $\chi^2$ with respect to
$\varphi_j$, $\Delta\varphi_j$, $I_j$ with $n = 3\times N +
1$ and the orbital parameters fixed) for several different
numbers of spots are shown in Figure 1 for the {\it Exosat} fits
and in Figure 2 for the {\it Ginga} fits.  Figure 3 shows how
the $\chi^2$ calculated from equation (2.1) depends on the
postulated number of spots. One concludes from these Figures
that the fits of the models with 3 or more spots are very good.
For {\it Exosat} data the best fit is given by the model with 4
spots, for {\it Ginga} the minimum of $\chi^2$ is very shallow
and one can only say that the best fit is given by models with
5, 6, or 7 spots.

Figure 4 and Table 1 show the locations, sizes and intensities
of the spots from the best fit models to the {\it Exosat} and {\it
Ginga} data. The basic argument of our paper is that the five
major spots have very similar phases. The maximal phase difference
between the corresponding {\it Exosat} and {\it Ginga} spots is
$\Delta\varphi = 0.029$ ({\it i.e.} 10 degrees). We argue that
this coincidence could not possibly occur by pure chance. The
simplest estimate of the probability that for two sets of $N$
ordered numbers between $0$ and $1$ the differences between
corresponding numbers are less than $\Delta \varphi$ is
$p_N(\Delta\varphi) \approx (\Delta \varphi)^N$. This gives
$p_5(0.029) = 2\times 10^{-8}$ and, if only the three major spots are
considered, $p_3(0.029) = 3\times10^{-5}$. Thus, one may conclude that the
coincidence between the three major spots in the {\it Exosat}
and {\it Ginga} light curves cannot occur by chance.

\smallskip
{\it ii) The probability of coincidence of two N-spot patterns.}
\smallskip

One can argue that the locations of the peaks of the flares in
the {\it Exosat} and the {\it Ginga} light curves are not truly
random, because if two flares are located too close to each
other, they would be mistaken for one spot. This means that
when the probability is calculated, the {\it Exosat} and {\it
Ginga} distributions of spots should be compared only with those
distributions in which all the spots in one set are distant
from each other by more than a given minimal separation $\delta
\varphi$. By increasing the minimal separation one increases the
probability of the coincidence of the two sets of numbers.
Thus, if in a real situation there are several flares separated
by $\delta \varphi_0$ or more, one {\it overestimates} the
probability by assuming $\delta \varphi = \delta \varphi_0$,
because these flares could be still recognized as separate with
a smaller separation constant.  Figure 4 shows that all the
spots in one set (either {\it Exosat} or {\it Ginga}) are
separated by more than $0.1$ in phase. Therefore, the data is
consistent with $\delta \varphi \simless 0.1$ (about 30 degrees).

We have calculated by Monte Carlo simulations what is the
probability $p_N(\delta \varphi,~\Delta
\varphi)$ that for two sets of $N$ spots separated by at least
$\delta \varphi$, the phase differences between the
corresponding spots are less than $\Delta \varphi$,
assuming in addition that the phases in the second set may all be adjusted by
a constant phase shift in order to make $\Delta \varphi$ as small as
possible. Results of
these calculations are presented in Figure 5.

{}From the fits discussed in this Section and presented in Figures
1, 2, 3, 4 and in Table 1 we obtained $3 \le N \le 5$, $\delta
\varphi \le 30$ degrees, and $\Delta \varphi \le 10$ degrees. As
can be seen from Figure 5, this corresponds to an upper
limit to the probability that such a coincidence could occur by
chance, $p_5 < 0.01$ (and $p_3 < 0.04$). Because the
probability is small, we conclude that the coincidence between
the phases of the peaks in the {\it Exosat} and {\it Ginga} light
curves is real and reflects an important intrinsic property of
the source.

The stability of the variability pattern found by us on the
basis of the {\it Exosat} and {\it Ginga} data and attributed to
{\it five} spots present on a fixed orbit, confirms an earlier
suggestion by Fiore {\it et al.} (1992a), made on the basis of
four {\it Exosat} observations, that the strong {\it fourth} harmonic
in the Fourier power spectrum of the variability of NGC 6814 may
be an intrinsic property of this source. The fourth harmonic
has its frequency equal to {\it five} times that of the fundamental
one.

Figure 6 shows the vectors corresponding to shifts from the {\it
Exosat} to {\it Ginga} positions of the spots. The Figure seems
to suggest that the spot pattern undergoes systematic rather
than chaotic changes. It would be very important to see whether
this behaviour is confirmed by the {\it Rosat} data because this
may be crucial in deciding the physical nature of the spots.

%%%%%%%%%%%%%%%%%%%%%%%%%%%%%%%%%%%%%%%%%%%%%%%%%%%%%%%%%%%%%%%%%%%%%%%%%%%%%%%
\vskip 0.4truecm
\centerline {\bf 3. Discussion}
\vskip 0.4truecm
%%%%%%%%%%%%%%%%%%%%%%%%%%%%%%%%%%%%%%%%%%%%%%%%%%%%%%%%%%%%%%%%%%%%%%%%%%%%%%%

Although the physical nature of the spots is mostly irrelevant for the
arguments presented here, we would like to point out that Abramowicz
{\it et al.} (1992a) discussed observational evidence in favour of the
possibility that the small bright spots could be small,
transient vortices. The strong spots responsible for the
periodic signal in NGC 6814 could be giant and long-lived
vortices, similar to the Jupiter's Great Red Spot. The Great Red
Spot is a very long lived vortex: it has survived more than
$3\times 10^5$ rotational periods of Jupiter since it was
discovered by Galileo. It is known that the Great Red Spot
slowly moves along its orbit (has a longitude motion).
Numerical hydrodynamical simulations (see Abramowicz {\it et
al.} 1992a for references) show that typically in a situation
where many interacting vortices are present, very strong
vortices are rare and have long lifetimes, while small vortices
are frequent and relatively short lived.

Other possible explanations for stable and strong bright spots
on the accretion disk surface include a stable complex of
magnetic spots such as in rapidly rotating solar type stars, a
perturbation due to an additional center of accretion, {\it
e.g.} a small star or black hole orbiting inside the accretion
disk, non-axially symmetric instabilities, spiral structures,
{\it etc.}

The value of the mass obtained from the fit, $M = 9\times 10^6 M_{\odot}$,
agrees very well with that estimated by
Padovani \& Rafanelli (1988) who got $M = 6.68 \times 10^6
M_{\odot}$ from a kinematic analysis of the the broad emission
lines. The same authors estimated the bolometric luminosity of
NGC 6814 to be about $10^{44}$ erg sec$^{-1}$, and therefore the
bolometric to Eddington luminosity ratio to be $\lambda \approx 10^{-1}$.

The high inclination obtained in our fit, $\theta = 85^o$, needs
a few words of
comment. Recently Yamauchi {\it et al.} (1992) analysed the {\it
Ginga} data on the strong X-ray reflection component consisting
of an iron line and broad hump of emission extending from about
10 keV to 100 keV, of which only the lower end is yet detected.
(See also Nandra {\it et al.} 1992 and Matsuoka, 1992.) Yamauchi
{\it et al.} calculated the width of the iron line profile
(FWHM) for different inclinations and concluded that the
inclination of the disk must be rather low, $\theta \approx
8^o$, because for higher inclinations the Doppler effect would
broaden the profile more than the observational upper limit $(0.4
\pm 0.4)$ keV obtained by Kunieda {\it et al.} (1990). However,
in our opinion this result strongly depends on the assumed
geometry.  According to our calculations, when one includes the
effect of occultation of the innermost part of the disk by its outer
parts (this is important at high inclination, but was ignored
by Yamauchi {\it et al.}), the observational limit for the FWHM
is met, and the shape of the light curve is only very little
changed. The presence of absorbtion features in NGC 6814 was noticed
by several authors,  {\it e.g.} by Done {\it et al.} (1992b),
and this independently points to a possible importance of occultations.

One should stress that arguments based on the iron
line cannot be at the present time considered as very reliable
because some of the
most fundamental aspects of the data on the iron line in NGC
6814 are not understood at all. In particular, the equivalent
width of the line, $ EV = 300 - 500$ eV, is 2 to 3 times larger
than expected from the standard theory. It is worth quoting here
a popular explanation for this which also assumes a high
inclination in agreement with what we have found: ``One possible
solution is that they [NGC 6814 and NGC 5548] are observed at
high inclination so that what would otherwise be secondary
effects now dominate.'' (Fabian 1992).

%%%%%%%%%%%%%%%%%%%%%%%%%%%%%%%%%%%%%%%%%%%%%%%%%%%%%%%%%%%%%%%%%%%%%%%%%%%%%%%
\vskip 0.4truecm
\centerline {\bf 4. Conclusions}
\vskip 0.4truecm
%%%%%%%%%%%%%%%%%%%%%%%%%%%%%%%%%%%%%%%%%%%%%%%%%%%%%%%%%%%%%%%%%%%%%%%%%%%%%%%

We found that the variability
patterns present in the {\it Exosat} and {\it Ginga} folded
light curves are remarkably similar --- the corresponding peaks
differ in phase by less than 10 degrees.
Predictions from the bright spots model agree very well, in a strict
quantitative way, with the stability of the variability pattern.
According to our interpretation of the data,
not only does one
clearly and directly see relativistic rotation of the disk in
the {\it Exosat} and {\it Ginga} folded light curves, but there
is also a serious possibility that these curves display, again
quantitatively, the influence of the slow inward accretion flow
which proceeds on the viscous timescale.

Our model is directly {\it testable}: it should be definitively rejected
if some future observations in a similar range of X--ray spectrum
show no periodic variability or a different variability pattern.
However, if the stability of the variability pattern suggested in our paper
is confirmed
by a third independent observation ({\it e.g.} in UV or optical, or
in X-rays by {\it Rosat}), then the probability that in all three cases the
coincidence occurs by pure chance would be negligibly small.
Then, not only our model will be proved correct, but in addition the
long awaited unquestionable proof of the correctness
of the AGN paradigm will be finally at hand.

The quality of the fits of the model light curves to those
observed is excellent, as we have demonstrated in terms of the
$\chi^2$ test.
Thus, we conclude that a detailed {\it quantitative} analysis of
the data strongly supports the bright spots
model. Whether the other models
(Syer, et al., 1991, Rees, 1992, Sikora and Begelman, 1992, Done and King,
1992)
could also explain
the variability pattern stability in the same accurate
and quantitative way remains to be seen.

We thank Aldo Treves for his stimulating critical remarks.

%%%%%%%%%%%%%%%%%%%%%%%%%%%%%%%%%%%%%%%%%%%%%%%%%%%%%%%%%%%%%%%%%%%%%%%%%%%%%%%
%                            F I G U R E S
%%%%%%%%%%%%%%%%%%%%%%%%%%%%%%%%%%%%%%%%%%%%%%%%%%%%%%%%%%%%%%%%%%%%%%%%%%%%%%%

\vfill
\eject
\centerline {\bf Figure and Table captions}
\vskip 0.4truecm
\par\noindent
{\bf Figure 1:} The best fits to the {\it Exosat} folded light curve
for the bright spot model with $N = 1, ~3, ~4,~5$ spots.  The
theoretical curves are shown by the broken lines. The {\it
Exosat} data was collected during the ``long look'' 1985/289
with the duration of about two days and corresponds to photon
energies from 2 keV to 6  keV.
\par\noindent
{\bf Figure 2:} The best fits to the {\it Ginga} folded light
curve for the bright spot model with $N = 4,~5,~6, ~7$ spots.
The theoretical curves are shown by the broken  lines. The {\it
Ginga} light curve corresponds to the total duration of about
three days and photon energies from 2.24 keV to 5.69 keV.
\par\noindent
{\bf Figure 3:} The $\chi^2$ values for the best fits to both
{\it Exosat} and {\it Ginga} folded light curves with different
number of spots.
\par\noindent
{\bf Figure 4:} The spot patterns used in our model for {\it
Exosat} and {\it Ginga} fits.
\par\noindent
{\bf Figure 5:} Probability $p_N(\delta \varphi,~\Delta
\varphi)$ that two sets of $N$ observed phases (ordered numbers
between $0$ and $1$ separated in each set by at least $\delta
\varphi$) coincide with accuracy at least $\Delta \varphi$. Both
$\delta \varphi$ and $\Delta \varphi$ are given in degrees.
\par\noindent
{\bf Figure 6:} Shifts in the spots locations from {\it Exosat}
to six years later {\it Ginga} positions. A constant phase shift
of $10^o$ was added to all {\it Ginga} phases to make our
argument more apparent. The Figure clearly shows that the
pattern undergoes a {\it systematic} change: there is almost
exactly linear relation between the phase shift and the phase.
In our model such a relation is expected as a direct consequence
of the very slow inward accretion flow in addition to the
circular motion. The spots are not moving on circles, but rather
on very tight spirals, which correspond to spiral flow lines in
the accretion disk. Thus, the spots are located at slightly
different radii and therefore they have slightly different
periods. These differences are too small to affect the measured
period stability, but sufficient to produce, in six years, the
very small shifts seen in Figure 6.  Quantitatively, because the
shift for the spot No. 2 is by $20^o$ greater than for the spot
No. 4, one calculates directly from the observational data
that the spot No. 2 moves $\Delta r \approx 10^{-6}r_0 $ closer
to the centre than the spot No. 4.  This number should be
approximately of the same order as the  ratio of the orbital to
viscous time scale, estimated from the standard accretion disk
theory, $t_{\rm vis}/t_{\rm orb} \approx \lambda^2 \alpha$.
Thus, the Shakura-Sunyaev $\alpha$ viscosity parameter
should be of the order of $10^{-4}$, which is rather low, but
quite reasonable.
\par\noindent
{\bf Table 1:} The three major spots are indicated by asterisk
$\ast$. Intrinsic intensities and energies of spots are given in
arbitrary units.
%%%%%%%%%%%%%%%%%%%%%%%%%%%%%%%%%%%%%%%%%%%%%%%%%%%%%%%%%%%%%%%%%%%%%%%%%%%%%
% %                          R E F E R E N C E S %
%%%%%%%%%%%%%%%%%%%%%%%%%%%%%%%%%%%%%%%%%%%%%%%%%%%%%%%%%%%%%%%%%%%%%%%%%%%%%
\vfill
\eject
\centerline {\bf References}
\vskip 0.4truecm

\ref Abramowicz M.A., 1992, in {\it
     Testing the AGN Paradigm}, eds. S.S. Holt, S.G. Neff, \& C.M.
     Urry, American Institute of Physics, New York

\ref Abramowicz M.A., Bao G., Lanza A., et al., 1989,
     in {\it Proceedings of the 23rd ESLAB Symposium on Two
     Topics in X-ray Astronomy}, eds. J. Hunt and B. Battric,
     ESA, SP-296

\ref Abramowicz M.A., Bao G., Lanza A., et al., 1991,
     {A\&A}, {\bf 245}, 454

\ref Abramowicz M.A., Lanza A., Spiegel E.A., et al., 1992a,
     {Nat}, {\bf 356}, 41

\ref Abramowicz M.A., Bao G., Fiore F., et al., 1992b, in
     {\it Physics of Active Galactic Nuclei}, eds. W.J. Dushl
     \& S.J. Wagner, Springer, Heidelberg

\ref Abramowicz M.A., Bao G., Fiore F., et al., 1992c,
     {Nat}, submitted

\ref Bao, G, 1992, {A\&A}, {\bf 257}, 594

\ref Bao, G,  Stuchl{\'\i}k, Z., 1992, {ApJ}, (in press)

\ref Done C., King A., 1992, {Nat}, (submitted)

\ref Done C., Madejski G.M., Mushotzky R.F.,  et al., 1992a in {\it
     Testing the AGN Paradigm}, eds. S.S. Holt, S.G. Neff, \& C.M.
     Urry, American Institute of Physics, New York

\ref Done C., Madejski G.M., Mushotzky R.F., et al.,
     1992b, {ApJ}, in press

\ref Dultzin-Hacyan D., Schuster W.J., Parraro L., et al.,
     1992, {AJ}, {\bf 103}, 1769

\ref Fabian A.C., 1992, in {\it Testing the AGN Paradigm},
     eds. S.S. Holt, S.G. Neff, \& C.M. Urry, American Institute
     of Physics, New York

\ref Fiore F., Massaro E., Barone P., 1992a, in {\it Physics of
     Active Galactic Nuclei}, eds. W.J. Dushl \& S.J. Wagner,
     Springer, Heidelberg

\ref Fiore F., Massaro E., Barone P., 1992b, {A\&A},
     {\bf 261}, 405

\ref Karas, V, Bao G., 1991, {A\&A}, {\bf 257}, 531

\ref Kunieda H., Turner T. J., Awaki H., et al.,  1990,
     {Nat}, {\bf 345}, 786

\ref Mastsuoka M., 1992, in {\it Testing the AGN Paradigm},
     eds. S.S. Holt, S.G. Neff, \& C.M. Urry, American Institute
     of Physics, New York

\ref Miller H.R., Noble J.C., Carini M.T., et al.,
     1992, in {\it Testing the AGN Paradigm},
     eds. S.S. Holt, S.G. Neff, \& C.M.  Urry, American
     Institute of Physics, New York

\ref Mittaz J.P.D., Branduardi-Raymont G., 1989, {MNRAS},
     {\bf 238}, 1029

\ref Nandra K., Pounds K.A., Steward G.C., 1992, in {\it
     Testing the AGN Paradigm}, eds. S.S. Holt, S.G. Neff, \& C.M.
     Urry, American Institute of Physics, New York

\ref Padovani P., Rafanelli P., 1998, {A\&A}, {\bf 205}, 53

\ref Rees M.J., 1992, in {\it The Renaissance of General Relativity
     \& Cosmology: A survey meeting to celebrate the 65th
     birthday of Dennis Sciama}, eds. G.F.R. Ellis, A. Lanza,
     J.C. Miller, Cambridge University Press, Cambridge

\ref Sikora M., Begelman M., 1992, {Nat}, {\bf 356},
     224

\ref Syer D., Clarke C.J., Rees M.J., 1991, {MNRAS},
     {\bf 250}, 505

\ref Veilleux S., Zheng W., 1991, {ApJ}, {\bf
     377}, 89

\ref Wallinder F.H., Kato S., Abramowicz M.A., 1992, {A\&AR},
     {\bf 4}, 79

\ref Witta P.J., Mangalam A.V., Chakrabarti S.K., 1992, in
     {\it Testing the AGN Paradigm}, eds. S.S. Holt, S.G. Neff,
     \& C.M.  Urry, American Institute of Physics, New York

\ref Witta P.J., Miller H.R., Carini M.T., et al., 1991,
     in {\it Structure and Emission properties of Accretion Disks},
     eds. C. Bertout {\it et al.}, Editions Fronti{\'e}res, Gif-sur-Yvette

\ref Yamauchi M., Matsuoka M., Kawai N., et al., 1992,
     {ApJ}, {\bf 395}, 453

\ref Zhang, X-H., Bao, G., 1991, {\it A\&A}, {\bf 246}, 21

\ref Zheng W., Veilleux S., Grandi M., 1991, {ApJ},
     {\bf 381}, 418

%%%%%%%%%%%%%%%%%%%%%%%%%%%%%%%%%%%%%%%%%%%%%%%%%%%%%%%%%%
\vfill\eject
\centerline{\bf Table 1.}
$$\vbox{\tabskip=0pt \offinterlineskip
\halign {\strut #&  		%column 1
\hfil #\hfil& 			%column 2
\hfil #\hfil&			%column 3
\hfil #\hfil&			%column 4
\hfil #\hfil&			%column 5
\hfil #\hfil&			%column 6
\hfil #\hfil&			%column 7
\hfil #\hfil&			%column 8
\hfil #\hfil&			%column 9
\hfil #\hfil&			%column 10
#
\cr
  Spot's&
  ~~Spot's phase\span &
  ~~Spot's width\span &
  Spot's intensity\span &
  ~~Spot's energy \span &
  Phase difference &
\cr
  number &
  ~~Ginga~~ &
  ~~Exosat~~&
  ~~Ginga~~&
  ~~Exosat~~&
  ~~Ginga~~&
  ~~Exosat~~ &
  ~~Ginga~~&
  ~~Exosat~~&
  Ginga--Exosat&

\cr \noalign{\hrule}
$^{\ast}1$ & 0.10 & 0.08 & 0.064 & 0.064 & 1.00 & 0.35 & 1.00 &
0.35 & ~0.022\cr
& ($~37^o$) & ($~29^o$) &&&&&&& ($~~8^o$)\cr
2& 0.25 & 0.21 & 0.024 & 0.008 & 0.50 & 0.36 & 0.20 & 0.05 & 0.032 \cr
& ($~90^o$) & ($~77^o$) &&&&&&& ($~11^o$)\cr
3 & 0.33 & & 0.021 & & 0.30 & & 0.10 & & \cr
& ($119^o$) & &&&&&&& \cr
$^{\ast}4$ & 0.48 & 0.51 & 0.068 & 0.057 & 0.77 & 1.50 & 0.73 &
1.33 & -0.029\cr
& ($172^o$) & ($182^o$) &&&&&&& ($-10^o$)\cr
$^{\ast}5$ & 0.82 & 0.84 & 0.032 & 0.032 & 1.20 & 0.30 & 0.60 &
0.15 & -0.016 \cr
& ($296^o$) & ($301^o$) &&&&&&& ($-~6^o$)\cr
6 & 0.96 & 0.94 & 0.016 & 0.008 & 0.36 & 0.16 & 0.10 &
0.10 & 0.019\cr
& ($344^o$) & ($337^o$) &&&&&&& ($~~7^o$)\cr
\noalign{\hrule}
}}$$
\bye